\documentstyle[aps,prd,eqsecnum]{revtex}
\input psfig.sty

\begin{document}

\title{Lyapunov timescales and black hole binaries}
\author{Neil J. Cornish${}^{*}$ and Janna Levin${}^{**}$}
\address{${}^{*}$ Department of Physics, Montana State University, Bozeman, MT 59717}
\address{${}^{**}$ DAMTP, Cambridge University,Wilberforce Rd., Cambridge CB3 0WA }

\twocolumn[\hsize\textwidth\columnwidth\hsize\csname
           @twocolumnfalse\endcsname

\maketitle
\widetext

\begin{abstract}

Black holes binaries support unstable orbits
at very close separations. In the simplest case of geodesics around
a Schwarzschild black hole the orbits, though
unstable, are regular. Under perturbation the unstable orbits
can become the locus of chaos.
All unstable orbits, whether regular or chaotic, can be quantified by
their Lyapunov exponents.
The exponents are
observationally relevant since the phase of gravitational
waves can decohere in a Lyapunov time.
If the timescale for dissipation due to gravitational waves is shorter
than the Lyapunov time, chaos will be damped and essentially unobservable.
We find the two timescales can be comparable.
We emphasize that the Lyapunov exponents must only be used
cautiously for several reasons: they are relative and depend on the
coordinate system used, 
they vary from orbit to orbit, and finally they can be deceptively
diluted by transient behaviour for orbits which pass in and out of
unstable regions.

\end{abstract}

\medskip
\noindent{04.30.Db,97.60.Lf,97.60.Jd,95.30.Sf,04.70.Bw,05.45}
\medskip
]

\narrowtext
 
\setcounter{section}{1}

Newtonian gravity predicts the elliptical planetary orbits 
around the sun which Kepler
described. Einstein gravity predicts precessing
elliptical orbits around a central star, thereby reconciling the 
precession of the perihelion of Mercury with general relativity.
In the extreme case of a central black hole, there are also 
a simple set of {\it unstable} circular orbits 
in addition to the usual stable circular orbits. 
Related to these are the homoclinic orbits which lie on the boundary
between dynamical stability and instability \cite{{bc},{loc}}.
To the list of possible orbits, a set of chaotic
orbits has recently been added for rapidly
spinning black holes \cite{{sm},{me1},{me2}} (and 
for the interesting but less physically realistic
Majumdar-Papapetrou black hole pairs of equal mass and charge
\cite{{bhs},{frank},{cf},{meglow}}).  
With the future gravitational wave experiments 
LIGO and LISA we hope to see these innermost orbits and 
reconstruct a map of spacetime around black holes.

The instability of the innermost orbits around black holes will etch
certain landmarks in a gravitational wave map.
Gravitational waveforms of neighboring orbits
will dechohere in a time scale set by the instability \cite{njc1}.
If the timescale for dissipation through gravitational radiation is faster
than the instability timescale, then chaos will be damped and the 
gravitational wave signal will not observably decohere.

The simple set of unstable circular orbits around a Schwarzschild
black hole are a consequence of the
nonlinearity of general relativity. 
Their instability can be quantified
by a positive Lyapunov exponent \cite{njc1}.
Although Lyapunov exponents are often associated with chaotic dynamics,
the geodesics around a Schwarzschild black hole are not chaotic:
the orbits are fully soluble and therefore integrable.
However,
under perturbation, chaos is likely to develop along the unstable
circular and homoclinic orbits.
An example of this has been found when
the black holes spin.
The nonlinearity degenerates to a nonintegrability and chaos
\cite{{sm},{me1},{me2}}. 
The number of unstable periodic orbits 
proliferates so that they have to pack themselves into a fractal
in order to crowd into that region of phase \cite{{frank},{me2}}.
The unstable orbits will have positive Lyapunov exponents 
\cite{{rs},{usl}} and will emerge as fractals in phase space \cite{{me1},{me2}}.

The Lyapunov exponents, while a seemingly useful tool, have uncomfortable
shortcomings in the context of general relativity.
Firstly, the Lyapunov exponents vary from orbit to orbit and so do not
have the surveying power to scan the collective
behaviour of all orbits that
fractals methods do.
Secondly, the Lyapunov exponents are a measure of the deviation of two
neighboring orbits in time and therefore overtly depend on the time coordinate
used. Since time is relative so too are the Lyapunov exponents.
The relativity of the Lyapunov exponents has been known to erroneously lead 
to zero Lyapunov exponents for truly
chaotic systems \cite{{barrow},{book},{mixm},{sk}}.
Importantly, topological measures of chaos such as 
fractals are coordinate invariant and are not plagued by the relativism
of space and time
\cite{{frank},{mixm}}.

In rare cases when there is a prefered time direction the ambiguity
of time can be avoided.
For the simplest case of a Schwarzschild black hole
there is a timelike Killing vector which selects a prefered time direction.
In other words, from our position asymptotically far away from the black
hole,
we use a well defined time coordinate in our observations. As long as we
conscientiously compare all timescales in the same coordinate system,
we should get meaningful comparisons.

We investigate the stability of three types of black hole binary:
(i) Schwarzschild black hole (non-spinning, test
particle motion), (ii) the Post-Newtonian (PN) expansion of the
two-body problem (non-spinning black holes), and 
(iii) the chaotic orbits of spinning black holes in the PN-expansion.

The stability analysis begins with
the equations of motion summarized as
\begin{equation}
\frac{d X_{i}}{dt} = H_{i}(X_{j}) \, .
\end{equation}
To analyze the stability of a given orbit we linearize the equations
of motion about that orbit
\begin{equation}
\frac{d\,  \delta \! X_{i}(t)}{dt} = K_{ij}(t)\,  \delta \! X_{j}(t) \, ,
\end{equation}
with
\begin{equation}
K_{ij}(t) = \left. \frac{\partial H_{i}}{\partial X_{j}}\right|_{X_{i}(t)}
\end{equation}
the linear stability matrix.
The solution to the linearized equations can be written as
\begin{equation}
\delta \! X_{i}(t) = L_{ij}(t)\, \delta\! X_{j}(0) \, 
\end{equation}
in terms of the evolution matrix which must obey
	\begin{equation}\label{evo}
	\dot L_{ij}(t)=K_{im} L_{mj}(t)\, 
	\end{equation}
and $L_{ij}(0) = \delta_{ij}$.
A determination of the eigenvalues of $L_{ij}$ 
leads to the principal Lyapunov exponent.
Specifically
	\begin{equation}\label{L}
	\lambda = \lim_{t \rightarrow \infty} \frac{1}{t} \log \left(
	\frac{ L_{jj} (t)}{L_{jj} (0)} \right) \, .
	\end{equation}

\section{Schwarzschild orbits}

\subsection{Circular orbits}

Here we evaluate the Lyapunov exponent of unstable
orbits around the Schwarzschild black hole. We consider the usual
geodesics of a non-spinning, light companion.
We work
in Schwarzschild time, the time measured by an observer asymptotically
far from the black hole. The Lyapunov exponent was already evaluated in
Ref.\ \cite{njc1} in a different time coordinate system. 
This exemplifies the ambiguity of time. 
Still, the timescales which were compared 
in that paper were all measured in the same coordinate system
and therefore the general conclusions of Ref.\ \cite{njc1} still hold.

To isolate $\lambda$,
we begin with the  Lagrangian for a (non-spinning) 
test particle in the Schwarzschild
spacetime 
\begin{eqnarray}
{\cal L} &=& \frac{1}{2}\left(- \frac{(r-2)}{r}\left(\frac{ d t}{ds}\right)^2
+\frac{r}{r-2}\left(\frac{ d r}{ds}\right)^2 \right. \nonumber \\
&& \quad \left. + r^2 \left(\frac{ d \theta}{ds}\right)^2
+r^2\sin^2\theta \left(\frac{ d \phi}{ds}\right)^2 \right).
\end{eqnarray}
The black hole mass has been set to unity.
We consider motion in an equatorial plane to eliminate the 
cylic $\theta$ variable by setting it equal to $\pi/2$ and
define the canonical momenta by $\delta {\cal L}/\delta(dq/ds)=p_q$:
	\begin{eqnarray}\label{p}
	-p_t &=& \frac{r-2}{r}\frac{dt}{ds} = E\nonumber \\
	p_\phi &=& {r^2}\frac{d\phi}{ds} = L \nonumber \\
	p_r &=& \frac{r}{r-2}\frac{dr}{ds} \, .
	\end{eqnarray}
To change into Schwarzschild time $t$, we use
eqn.\ (\ref{p}) to define the transformation
	\begin{equation}
	\frac{d}{ds}=\frac{Er}{r-2}\frac{d}{dt} \, .
	\end{equation}
Notice that in Ref.\ \cite{njc1} an unusual time coordinate $t^\prime$
was used instead which was defined by the transformation
	\begin{equation}
	\frac{d}{ds}=\frac{r-2}{E r}\frac{d}{dt^\prime} \, .
	\end{equation}
We will redo the stability analysis in Schwarzschild time $t$.
The equations of motion can be derived through
\begin{equation}
	\frac{\delta{\cal L}}{\delta dr/ds}-\frac{\delta {\cal L}}
{\delta r}=0 \, 
	\end{equation}
and reduce to a two-dimensional system:
	\begin{eqnarray}
	\dot p_r &=& -\frac{E}{r(r-2)}-\frac{r-2}{r^3}\frac{p_r^2}{E}
		+\frac{r-2}{r^4}\frac{L^2}{E}\nonumber \\
	\dot r &=& \left (\frac{r-2}{r}\right )^2\frac{p_r}{E} 
	\end{eqnarray}
where an overdot denotes differentiation with respect 
to Schwarzschild time $t$.
To compare with Ref.\ \cite{njc1}, we first consider circular orbits.
Linearizing the equations of motion 
with $X_i(t)=(p_r, r)$ about orbits of constant $r$ gives
\begin{equation}\label{kcirc}
K_{ij} = \pmatrix{0 & \frac{2(r-1)E}{r^2(r-2)^2}-\frac{L^2}{E}\frac{(3r-8)}{r^5}\cr
\left (\frac{r-2}{r}\right )^2\frac{1}{E}  &
0
}
\end{equation}
The eigenvalues along circular orbits are
	\begin{eqnarray}\label{lamcirc}
	\lambda_{\pm}= \pm 
	 \left [\frac{2(r-1)}{r^4}
	-\frac{(3r-8)(r-2)^2}{r^7}\frac{L^2}{E^2}\right ]^{1/2}
	\, .
	\end{eqnarray}

For the unstable circular orbit at $r=4$ the angular momentum 
and energy are $L=4,E=1$ respectively and the eigenvalues of
$K_{ij}$ are
	\begin{equation}\label{lpm}
	\lambda_{\pm}=\pm\frac{1}{8\sqrt{2}} \, .
	\end{equation}
The conservation of energy ensures that in these canonical
coordinates, the Lyapunov exponents must come in plus-minus pairs
to conserve the volume of phase space.

The unit normalized eigenvectors corresponding to $\lambda_\pm$ are
	\begin{eqnarray}
	{\bf e}_+ &=&\frac{1}{3} \left(-1, 2\sqrt{2}\right) \nonumber \\
	{\bf e}_- &=&\frac{1}{3} \left(1, 2\sqrt{2}\right) \, .
	\end{eqnarray}
In this eigenbasis $K_{ij}$ is diagonal with
	\begin{equation}
	K_{ij}=\pmatrix{\frac{1}{8\sqrt{2}} & 0 \cr
	0 & -\frac{1}{8\sqrt{2}} 
	}
	\end{equation}
so that
	\begin{equation}
	L_{ij}=\pmatrix{ \exp(\frac{1}{8\sqrt{2}}t) & 0 \cr
	0 & \exp(-\frac{1}{8\sqrt{2}}t) 
	}\, .
	\end{equation}
Notice that in Ref.\ \cite{njc1}, the analysis was carried out in a 
four-dimensional coordinate system: $X^\prime_i=(p_r,p_\phi,r,\phi)$.
The additional coordinates are unimportant in the dynamical study and 
only hampered the diagonalization of $K_{ij}$ \cite{njc1}. If we 
were to redo the stability analysis in three coordinates
$X^{\prime \prime}=(p_r,p_\phi,r)$, we would get the same eigenvalues
(\ref{lpm}) and an additional $\lambda=0$. If we move up to the four
coordinates of $X_i^\prime$ we add yet another $\lambda=0$ giving
a degenerate set of eigenvalues. The matrix $K_{ij}$ cannot be
diagonalized in the event of degenerate eigenvalues which leads to
an unnecessary complication. For this reason, we stick to the pertinent
two-dimensional system $X_i=(p_r,r)$.

The relativity of time and the influence on the Lyapunov exponent is 
apparent at this stage. In Schwarschild time $t$ we find
$\lambda_{\pm}=\pm\frac{1}{8\sqrt{2}}$ while in the time $t^\prime$
used in Ref.\ \cite{njc1}, at $r=4$ the exponents were
found to be $\lambda^\prime_{\pm}=\pm \frac{1}{2\sqrt{2}}$. The Lyapunov
exponents are not coordinate invariant. However at $r=4$, 
$t^\prime = t/4$ and it follows that the combination
	\begin{equation}
	\lambda t = \lambda^\prime t^\prime \, 
	\end{equation}
is invariant.

We compare the Lyapunov timescale $T_\lambda=1/\lambda$
to the gravitational wave timescale $T_w=2\pi/\dot \phi$.
For the orbit at $r=4$, 
$T_\lambda/T_w=\sqrt{2}/\pi \approx 0.45$. The Lyapunov timescale
is less than about one orbit around the central black hole.
Notice that even though Ref.\ \cite{njc1} operated in an unusual
time, the calculations were performed self-consistently so that
the ratio of $T_\lambda$ to $T_w$ is correct.
The Lyapunov timescale is shorter than the gravitational wave timescale,
making the instability observationally relevant.

In principle we could also compare $T_\lambda$ to the decay time due to 
energy lost in the form of gravitational radiation. For a test-particle
in a circular orbit around a Schwarzschild black hole the decay time is
$T_d=(5/256)r^4/\mu$ where $\mu$ is the reduced mass. At $r=4$ this is
$T_d=5/(3\mu)\gg 1$ since in the test-particle limit $\mu \ll 1$
and $T_\lambda $ will be shorter than the decay time, again making the
instability observationally relevant.

\subsection{Homoclinic orbits}

Although the unstable circular obits are often emphasized
they are actually a subset
of the pertinent orbits. The division between stability and instability
for a Schwarzschild black hole is often taken to be the innermost
stable circular orbit (ISCO). The ISCO is actually the saddle point
at which the unstable circular orbit coincides with the stable circular
orbit. From a dynamical systems point of view, the true division 
between stability and instability occurs more generally
along the homoclinic orbits \cite{{bc},{loc}}.

\begin{figure}
\centerline{\psfig{file=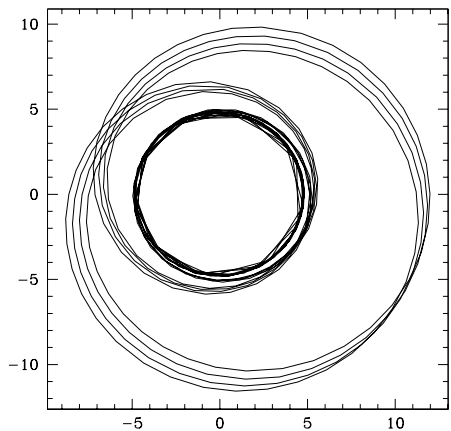,width=2.5in}}
\caption{A segment of the homoclinic orbit with $\beta=1/4$.
\label{orbit}}  \end{figure}

The stable manifold of a periodic orbit is defined as 
the set of points in phase
space that, when evolved forwards in time, approach the periodic orbit. 
The unstable manifold is the set of
points in phase that, when evolved backwards in time, approach the periodic 
orbit.
For an integrable, nonchaotic system,
the stable and unstable manifolds can intersect each other
along a single orbit. This orbit is called homoclinic if it 
approaches the same fixed point in the past and in the future.
For black holes the 
homoclinic orbits begin at an unstable 
circular orbit, roll out to a maximum radius
and fall back in to the same unstable circular orbit.
An example is shown in fig. \ref{orbit}.
The homoclinic orbits are sometimes called zoom-and-whirl orbits in 
the gravitational wave literature because the orbits whirl around the
center of mass and then zoom out into an ellipse before whirling in again.

Under perturbation the homoclinic orbits can become the site of
a homoclinic tangle. The tangle occurs when the stable and unstable
manifolds intersect transversely at an infinite number of points.
The intersection will no longer occur along a simple line but will
instead define a fractal set of chaotic orbits. In this section, we 
study the stability of the nonchaotic, simple set of homoclinic orbits
around a Schwarzschild black hole. In \S \ref{chaos} we study chaotic
orbits of spinning black holes.

As is well known, 
orbital motion around a Schwarzschild black hole conveniently reduces
to one-dimensional motion in an effective
potential
	\begin{equation}\label{cons}
	\frac{1}{2}\dot r^2 + V_{\rm eff}(r)=E
	\end{equation}
with
	\begin{equation}\label{eff}
	V_{\rm eff}(r)=E+\frac{(r-2)^3}{2E^2r^3}\left 
	({1+\frac{L^2}{r^2}
	}\right )-\frac{(r-2)^2}{2r^2}\, .
	\end{equation}
Circular orbits are solutions of $V_{\rm eff}=E$. For large enough
angular momentum there are two circular orbits, one unstable and one 
stable. As found in Ref.\ \cite{bc}, the homoclinic orbits have $E<1$ and
are described by the solution
	\begin{eqnarray}\label{homo}
	p_r &=& \pm \frac{r}{r-2}\left [ E^2-\frac{r-2}{r}\left (1+\frac{L^2}{r^2}\right )\right ]^{1/2} \nonumber \\
	\frac{1}{r} & = & \frac{1-2\beta}{6} + \frac{\beta}{2} \tanh^2(\sqrt{\beta} \phi/2)  \nonumber \\
	\end{eqnarray}
where
$0 \le \beta \le 1/2$ and $t(\phi)$ is a complicated function \cite{bc}.
\footnote{There appears to be typo in eqn. (1.5) of \cite{loc}.
Eqn. (\ref{homo}) has a factor 1/2 which is missing from the second term 
in eqn. (1.5).}
The circular orbits can also be parameterzied by $\beta$ as
	\begin{eqnarray}
	\label{circ}
	 r_{\rm unstable}&=&6/(1+\beta)	\nonumber \cr
	r_{\rm stable}&=&6/(1-\beta) 	\nonumber \cr
	L&=&2\sqrt{3/(1-\beta^2)} \nonumber \cr
	E&=&\frac{2-\beta}{3}\sqrt{2/(1-\beta)}\, .
	\end{eqnarray}

\begin{figure}
\centerline{\psfig{file=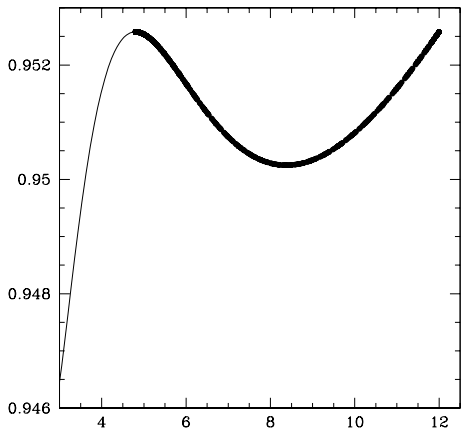,width=2.5in}}
\caption{The effective potential for orbits with 
$\beta=1/4$. The segment in bold marks the corresponding homoclinic trajectory.
The homoclinic orbit begins at the unstable circular orbit at the top 
of the hill ($r=4.8$), rolls out past the stable circular orbit in the valley
and on out to the maximum radius ($r=12$) before rolling
in again and climbing back up 
to the unstable orbit.
\label{veff}}  \end{figure} 

A homoclinic oribt starts at $r_{\rm unstable}$ and rolls out to 
	\begin{equation}\label{max}
	r_{\rm max}=6/(1-2\beta)\, 
	\end{equation}
winding around the black hole as it does so and then 
drops back in to $r_{\rm unstable}$. The ISCO is a homoclinic orbit
with $\beta=0$.  The first homoclinic
orbit at $\beta=1/2$ starts at the unstable orbit
at $r=4$ and rolls out to infinity before returning.
For $\beta=1/4$, $r_{\rm unstable}=4.8$, $r_{\rm stable}=8$ and
$r_{\rm max}=12$. 
The effective potential for the homoclinic orbit 
is drawn in fig.\ \ref{veff}. A segment of the orbit in the equatorial
plane is represented in fig.\ \ref{orbit}.

\begin{figure}
\centerline{\psfig{file=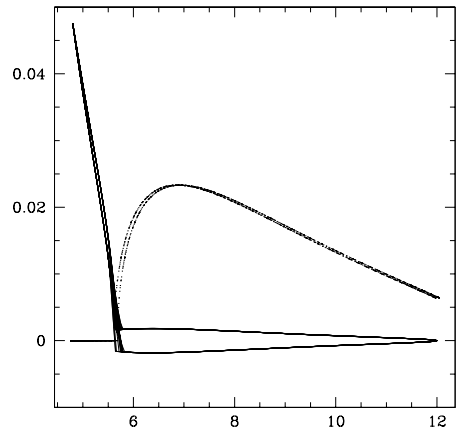,width=2.5in}}
\caption{$\beta=1/4$ The solid line is the Real part of the 
positive stability exponent and the dotted line is the Imaginary part.
\label{beta.25}}  \end{figure}

To analyze the stability we linearize to find
\begin{equation}\nonumber
K_{ij} = \pmatrix{
-\frac{2(r-2)}{r^3}\frac{p_r}{E} & \frac{2(r-1)E}{r^2(r-2)^2}+\frac{2(r-3)}{r^4}\frac{p_r^2}{E} -\frac{L^2}{E}\frac{(3r-8)}{r^5}\cr
\left (\frac{r-2}{r}\right )^2\frac{1}{E}  &
\frac{4(r-2)}{r^3}\frac{p_r}{E}  
}\, .
\end{equation}
The most general eigenvalues are
	\begin{eqnarray}\label{leg}
	\ell_{\pm}=\left (\frac{r-2}{r^2}\right )\frac{p_r}{E}
	& \pm &
	 \left [(2r-5)\frac{(r-2)^2}{r^6}\frac{p_r^2}{E^2} \right. \cr
&+& \left. 2\frac{(r-1)}{r^4}
	-\frac{(3r-8)(r-2)^2}{r^7}\frac{L^2}{E^2}\right ]^{1/2}
	\end{eqnarray}
Strictly speaking, $\ell$ is a stability exponent and is not identical
to the time averaged Lyapunov exponent defined in eqn. (\ref{L}).
Figure \ref{beta.25} shows the real and imaginary
parts of the positive stability exponent. As expected, the exponent
is postive near the unstable inner radius and becomes imaginary in the
vicinity of the stable circular radius dropping down to nearly zero as
it reaches the apihelion and then runs back through these values as it
moves back in to perihelion.

Because of this time variability in the stability along the orbit,
we have to be cautious in interpreting the timescales. 
The gravitational wave frequency will jag up and down as the orbit
zooms and whirls \cite{loc}. And it isn't obvious which timescales
to compare. Instead of using the variable, analytic result we could try 
a time average.

\begin{figure}
\centerline{\psfig{file=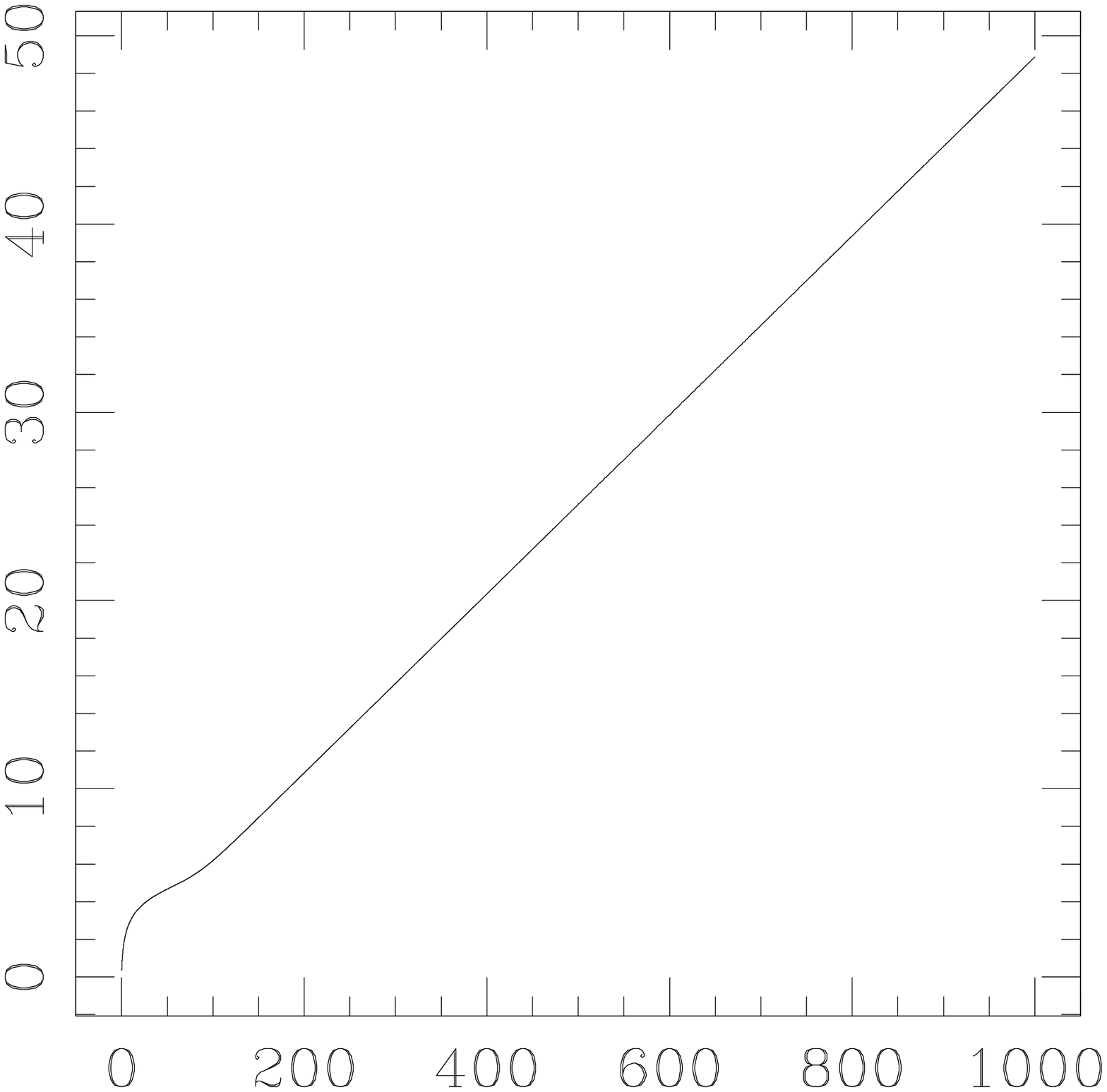,width=2.in}}
\centerline{\psfig{file=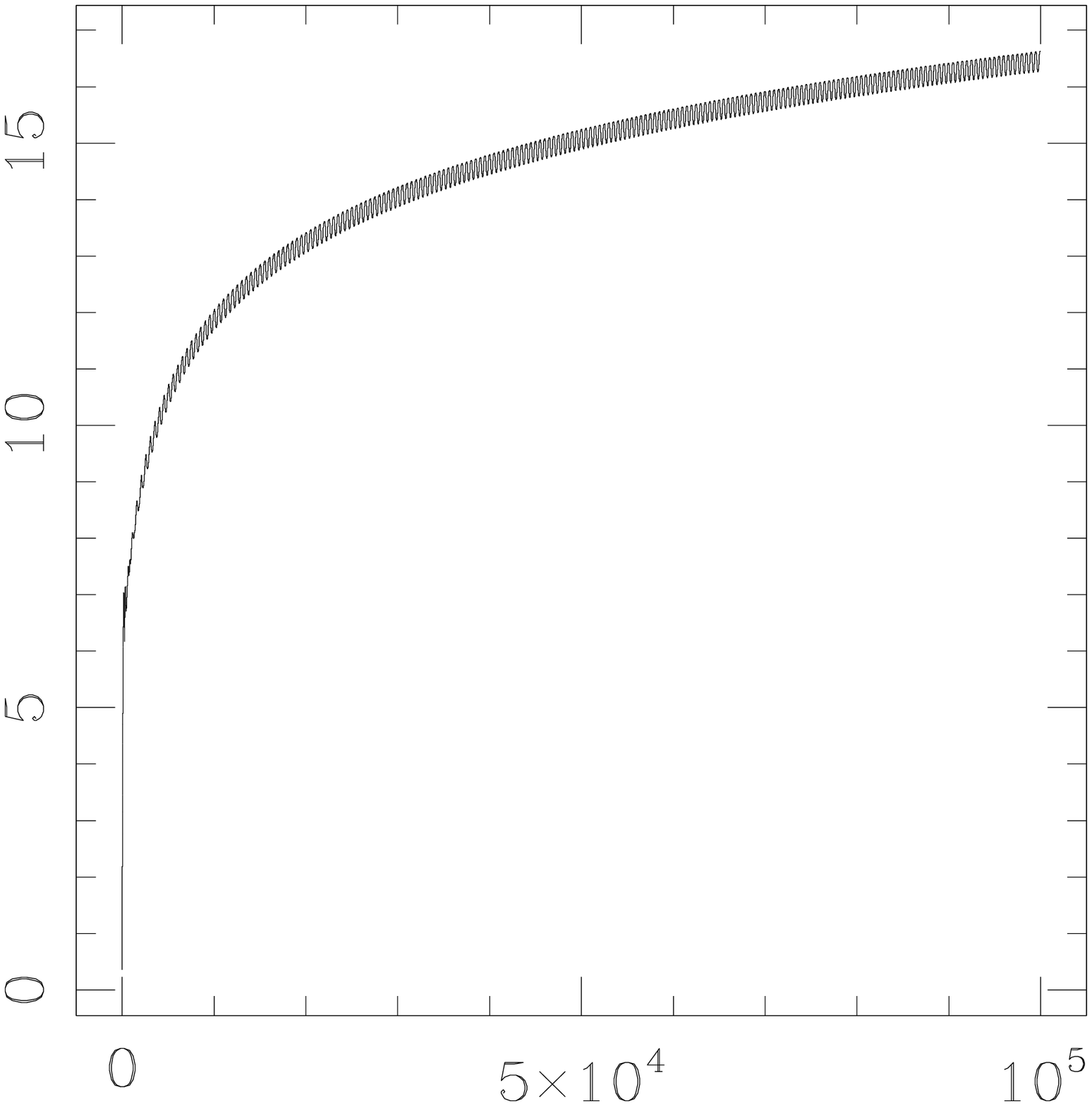,width=2.in}}
\caption{$\beta=1/4$. The slope of the line in the top panel is 
the numerically calculated Lyapunov exponent $\lambda \approx 0.475$
for the unstable circular orbit at $r=4.8$ using eqn.\ (\ref{L}). 
This matches exactly the analytic value predicted from 
eqn.\ (\ref{lamcirc}).
The slope of the line in the lower panel shows
zero Lyapunov exponent as calculated by eqn.\ (\ref{L}) for the homoclinic
orbit. This is to be contrasted with the analytic value of 
the stability exponent shown in fig.\ \ref{beta.25} which has positive segments
\label{le_b.25}}  \end{figure}

To this end we compare the analytic value of $\ell$ from eqn.\ (\ref{leg})
to the
time average Lyapunov exponent defined from eqn.\ (\ref{L})
	\begin{equation}
	\lambda = \lim_{t \rightarrow \infty} \frac{1}{t} \log \left(
	\frac{ L_{jj} (t)}{L_{jj} (0)} \right) \, .
	\end{equation}
For comparison we first look along the unstable circular orbit at $r=4.8$.
The upper panel of 
figure \ref{le_b.25} shows the Lyapunov exponent as defined by eqn.\ (\ref{L})
for the unstable
circular orbit. The exponent was obtained
by numerically integrating the $L_{ij}$ using eqn.\ (\ref{evo}) and 
figure \ref{le_b.25}
shows $\lambda t$ versus $t$ from eqn.\ (\ref{L}).
The numerically calculated value shown
in figure \ref{le_b.25} is identical to the analytic value given by 
eqn.\ (\ref{lamcirc}) of $\lambda \approx 0.475$.

However for a homoclinic orbit which begins at $r=4.8$,
the time averaged (\ref{L}) behaves as though there is no 
instability (lower panel of figure \ref{le_b.25}), 
when we know from the analytic result shown in fig.\ \ref{beta.25}
that there is.
The time averaged Lyapunov exponent
will vanish along this orbit even though it clearly has an unstable 
segment. 
We have to be very cautious therefore when we interpret the Lyapunov
exponent. 

For emphasis, if one was just scanning numerically, the mistaken
conclusion could be drawn that these orbits were dynamically simple.
This may turn out to be particularly important for the chaotic orbits
of \S \ref{chaos}.

\section{Post-Newtonian Orbits}

To move beyond the test particle limit, the two-body problem
has been expanded in a Post-Newtonian (PN) expansion approximation
to the fully relativistic two-body problem \cite{{kww},{linw},{extra1}}.
In this section we consider two black holes
which are not spinning.
In Ref.\ \cite{loc} the stability of the fixed points
in the PN equations to second-order (2PN) was tested
following \cite{kww}. 
We quote the results
of Ref.\ \cite{loc} here.
To second-order in the PN expansion,
the center of mass equations of motion for the binary orbit can be
written in harmonic coordinates as \cite{{kww},{linw},{extra1}}
	\begin{eqnarray}
	\ddot r_h &=&r_h \dot \phi^2-{1\over r_h^2}\left (A+B\dot
	r_h\right ) \label{eom1}\\
	\ddot \phi &=& -\dot \phi \left ({1\over r_h^2}B+2{\dot r_h
	\over r_h}\right )\label{eom2}
	\end{eqnarray}
where $M=1$ is the total mass of the pair.
The transformation between harmonic coordinates and Schwarzschild coordinates
is
$r_h =r -m$.  The form of $A(r_h,\dot r_h, \dot \phi)$ and $B(r_h,\dot r_h, \dot \phi)$
depends on the relative masses of the two black holes and on
the order of the PN expansion and can be found in 
Ref.\ \cite{kww}. 

\begin{figure}
\centerline{\psfig{file=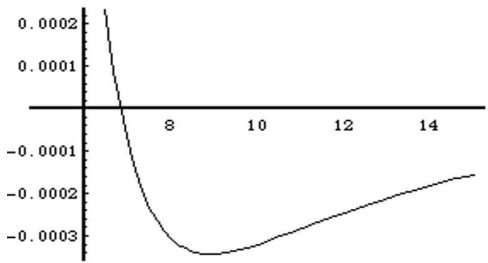,angle=0,width=2.5in}}
\caption{$\ell^2$ as a function of the constant circular
radius in harmonic coordinates, $r_{ho}$. $\ell^2>0$ corresponds to
unstable circular orbits and $\ell^2<0$ corresponds to
stable circular orbits.
\label{le_2pn}}  \end{figure}

As in Ref.\ \cite{kww}, 
the stability of the fixed points is tested by perturbing eqns.\
(\ref{eom1})-(\ref{eom2})
about a circular orbit to obtain,
	\begin{equation}
	K_{ij}
	=\pmatrix{0 & 1 & 0 \cr
	a & 0 & b \cr
	0 & c & 0}
	\label{mat}
	\end{equation}
with
	\begin{eqnarray}
	a &=&3\dot \phi_o^2-{m\over r_{ho}^2}
	\left ({\partial A\over \partial r_h}\right )_o  \nonumber \\
	b&=&2r_{ho}\dot \phi_o -{m\over r_{ho}^2}\left (	
	\partial A\over \partial \dot \phi\right )_o
	\nonumber \\
	c&=&-\dot \phi_o
	\left ({2\over r_{ho}} +{m\over r_{ho}^2}
	\left (\partial B\over \partial \dot r_h\right )_o\right )
	\, ,
	\label{abc}
	\end{eqnarray}
where $r_{ho}$ is the radius of the circular orbit in harmonic coordinates
and $\dot \phi^2_o=mA_o/r^3_{ho}$ is a function of the radius of the orbit
and is found explicitly in Ref.\ \cite{loc}.
The eigenvalues of (\ref{mat}) are,
	\begin{equation}
	\ell=0,\quad \ell_\pm=\pm(a+bc)^{1/2} \, .
	\end{equation}
Stable oscillations about a circular orbit correspond to imaginary
$\ell$ so that $a+bc<0$. 
Unstable orbits correspond to real positive $\ell$ and so have $a+bc>0$. 
The value of $\ell^2$ for equal mass binaries as a function 
of the circular radius is plotted in fig.\ \ref{le_2pn}.

Using the results of Ref.\ \cite{loc} we deduce that
in the test-mass limit,
the innermost {\it unstable}
circular orbit occurs at Schwarzschild radius 
$r_o=r_{ho}+1=4.96$. A comparison 
to the gravitational wave timescale in the PN expansion then gives 
$T_\lambda/T_w\approx 0.158$.
In the opposite extreme of equal mass binaries,
the innermost {\it unstable}
circular orbit occurs at $r_o=r_{ho}+1=5.78$ and
$T_\lambda/T_w\approx 0.21$.
A direct comparison to the Schwarzschild case isn't that
valuable. What is noteworthy is that the Lyapunov timescales
are again less than about one orbit around the center of mass.
Consequently, one expects the decoherence of the gravitational
waveform to be observationally significant. The instability
timescales are comparable to the decay times although of course
even a small loss in energy can induce merger for such an unstable
orbit.

The analysis extended to homoclinic orbits can be gleaned from 
Ref.\ \cite{loc}. 
The homoclinic orbits to 2PN order 
show similar features to the homoclinic
orbits of the Schwarzschild spacetime. The analytic Lyapunov exponent
will pass from positive to imaginary values as the orbit winds around
the center of mass. However, the time-averaged exponent will dilute 
these critical features.

None of these orbits are chaotic although they are unstable.
We turn to the chaotic orbits of rapidly spinning binaries next.

\section{Chaotic orbits}\label{chaos}

The dynamics can become chaotic when the homoclinic orbit is perturbed
leading to a homoclinic tangle. The intersection of the stable and unstable
manifold will no longer occur along a line but will intersect transversally
an infinite number of times. The fractal set of unstable chaotic
orbits lies along this tangled intersection. Ref.\ \cite{bc} studied generic
gravitational perturbations along the homoclinic orbits and found,
as they expected, that the dynamics could become chaotic. The physical
significance of the perturbations however wasn't clear and therefore
the observational consequences were difficult to assess. 

\begin{figure}
\centerline{\psfig{file=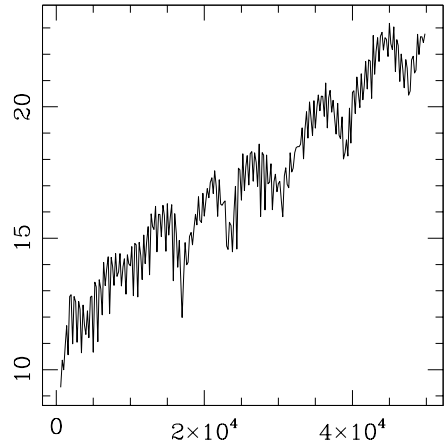,width=2.5in}}
\caption{The slope of the line is 
the Lyapunov exponent for the chaotic orbit of fig.\ \ref{orb_pns}.
\label{leL_pns}}  \end{figure} 

The chaotic dynamics discovered in 
Ref.\ \cite{sm} for a supra-maximally spinning test-particle
and for rapidly spinning pairs in the Post-Newtonian expansion
\cite{{me1},{me2}}, may occur along these homoclinic orbits. At the least
the chaotic behaviour kicks up most conspicuously in the vicinity of
these orbits.

\begin{figure}
\centerline{\psfig{file=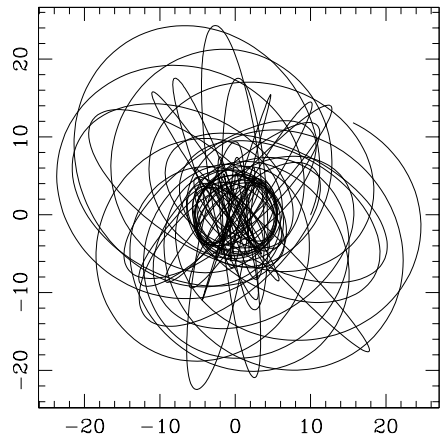,width=2.5in}}
\caption{A projection onto the plane 
of two maximally spinning black holes in a chaotic orbit.
\label{orb_pns}}  \end{figure}

We cannot determine the Lyapunov exponents analytically since the orbits
are not analytically soluble - the very meaning of nonintegrability.
We can however use eqn.\ (\ref{L}) to numerically determine the exponents
as we have done in Ref.\ \cite{usl}.
The Lyapunov exponent for a maximally spinning pair of black holes
is shown in fig.\ \ref{leL_pns}.
The value read from this 
is $T_\lambda \approx 11$
in units of windings around the center of mass. 
A segment of the orbit projected onto a plane
is shown in fig.\ \ref{orb_pns}.

For such an erratic orbit, it is not simple to define the 
gravitational wave timescale or the
radiation
reaction timescale. By starting the numerical simulation at a radius 
greater than 20 we found a rough estimate of $T_d\sim 4-5$ orbits
so that $T_\lambda/T_d > 2$ and the Lyapunov timescale is longer than
the dissipation timescale. For this 
orbit the gravitational waveform will not have time to decohere before
plunge and observations will not be severly disrupted by irregularity.
However a factor of 2 is a tight margin especially since the PN-expansion
is being pushed to extremes at such small separations.
 
It is important to stress that this is but one specific orbit and the ratio
of $T_\lambda/T_w$ will vary from orbit to orbit. Some chaotic orbits 
will undoubtedly have a longer Lyapunov timescale to dissipation timescale.
A broad survey of the irregular region of phase space would be valuable.
At 2PN order this is not so useful since the PN-expansion converges very
slowly to the full relativistic problem. Notice for instance that 
$T_\lambda$ is longer for this two-body approximation than it is for
the fully relativistic Schwarzschild orbits. 
This may be a consequence
of the slow convergence of the expansion and may show an underestimate
of the instability.
It is also possible that, like the homoclinic orbits, the measure of
instability is diluted by the time average across the orbit.
A survey at higher orders
may be useful but requires a greater than 3PN-expansion
that includes all spin arrangements and is not restricted to circular or 
quasi-circular orbits. We hope these higher orders will be available 
imminently and a survey of orbits will be viable in the near future.

\section*{Summary}

The Lyapunov exponents can be very useful for comparisons of physical
scales. However, Lyapunov exponents also have some shortcomings which
require we tread cautiously:

\noindent $\bullet $ Lyapunov exponents are relative. They depend on the
worldline of the observer and the time they measure.\par
\noindent $\bullet$ They vary from orbit to orbit and may not contain
generic information.\par
\noindent $\bullet$ They can give zero when averaged over orbits which move
in and out of unstable regions.\par
\noindent Therefore, while important and useful, the Lyapunov exponents
can be misleading and can only be used cautiously.

As far as we can trust them, the Lyapunov exponents
give an estimate of the importance of instability to observations
of gravitational waves. If the Lyapunov timescale is short compared
to the inverse 
frequency of the gravitational waves emitted and is short compared
to the
dissipation timescale then instability will cause an
observable decoherence of gravitational waves.
We found that \par
\noindent $\bullet$ the Lyapunov timescale
is shorter than both the gravitational wave timescale and the
dissipation timescale for unstable circular orbits in the 
approximation of a test-particle around a Schwarzschild black hole,\par
\noindent $\bullet$ the Lyapunov timescale
is shorter than the gravitational wave timesecale and comparable to the
dissipation timescale for unstable circular orbits in the 
2PN approximation in the absence of spins, \par
\noindent $\bullet $
and that the Lyapunov time was about a factor of 2 larger than the decay 
time for one
randomly sampled chaotic orbit of a pair of maximally spinning black 
holes in the 2PN expansion. 

The longer Lyapunov time for orbits in the 2PN approximation versus the 
test-particle approximation may be a real effect or 
it may be due to the slow convergence of the 2PN expansion to the full
nonlinear problem or finally it may be due to the time average over such 
a varied orbit.
In short, dissipation due to gravitational waves does abate chaos 
although the competition between chaos and dissipation is close.
Better approximations to the two-body problem
are needed to determine conclusively 
if chaos will affect observations of gravitational waves.

\section*{Acknowledgements}
NJC is supported in part by National Science Foundation Grant No. PHY-0099532.
JL is supported by a PPARC Advanced Fellowship and an award from NESTA.

\end{document}